\documentclass[twocolumn,showpacs,preprintnumbers,amsmath,amssymb]{revtex4}

\usepackage{graphicx}
\usepackage{dcolumn}
\usepackage{bm}

\begin{document}

\title{Electronic Structure and Fermiology of PuCoGa$_5$}

\author{Takahiro Maehira$^1$, Takashi Hotta$^1$, Kazuo Ueda$^{2,1}$,
and Akira Hasegawa$^3$}

\affiliation{
$^1$Advanced Science Research Center,
Japan Atomic Energy Research Institute,
Tokai, Ibaraki 319-1195, Japan \\ 
$^2$Institute for Solid State Physics,
University of Tokyo, Kashiwa, Chiba 277-8581, Japan \\
$^3$Niigata University, Niigata, Niigata 950-2181, Japan
}

\date{\today}

\begin{abstract}
By using a relativistic linear augmented-plane-wave method, we clarify
energy band structures and Fermi surfaces of recently discovered
plutonium-based superconductor PuCoGa$_5$.
We find several cylindrical sheets of Fermi surfaces with large volume,
very similar to CeMIn$_5$ (M=Ir and Co) isostructural with PuCoGa$_5$,
in spite of different $f$-electron numbers
between Ce$^{3+}$ and Pu$^{3+}$ ions.
The similarity is understood by a concept of electron-hole conversion
in a tight binding model constructed based on the $j$-$j$ coupling scheme.
Based on the present results, we provide a possible scenario to explain why
a transition temperature is so high as 18.5K in PuCoGa$_5$.
\end{abstract}



\maketitle


Recently it has been discovered that PuCoGa$_5$ exhibits superconductivity
\cite{Sarrao}.
Surprisingly its superconducting transition temperature
$T_{\rm c}$ is 18.5K, which is the highest among those yet
observed $f$-electron materials and high enough even compared with other
well-known intermetallic compounds.
It has been also found that PuRhGa$_5$ becomes superconducting
with $T_{\rm c}$=8.6K \cite{Wastin}.
These plutonium intermetallic compounds PuMGa$_5$ have the same HoCoGa$_5$-type
tetragonal structure as CeMIn$_5$, a family of cerium-based heavy fermion
superconductors \cite{Ce115}.
Note, however, that superconductivity occurs for M=Ir ($T_{\rm c}$=0.4K)
and Co (2.3K) in CeMIn$_5$, while antiferromagnetic (AFM) phase
has been found for M=Rh at ambient pressure.
Another isostructural material including uranium is UMGa$_5$ \cite{U115},
but superconductivity has not been found yet.
These HoCoGa$_5$-type compounds are frequently referred to as ``115''.

Regarding superconducting mechanism in the 115 compounds,
first let us consider Ce-115.
It has been widely considered that it is unconventional
$d$-wave superconductor induced by AFM spin fluctuations.
In fact, there are some evidences such as $T^3$ behavior in nuclear
relaxation rate \cite{NMR} and node structure measured by thermal
conductivity \cite{Izawa}.
For the phase diagram of Ce(Co,Rh,Ir)In$_5$ \cite{Pagliuso},
AFM phase is found to exist in adjacent to the superconducting phase.
Those experimental facts remind us of high-$T_{\rm c}$ cuprates,
but a clear difference from cuprates should be remarked.
Namely, high-$T_{\rm c}$ superconductivity in cuprates occurs by hole doping
into AFM insulators, while in Ce-115, $no$ hole doping is needed.
To understand the appearance of superconductivity induced by
AFM spin fluctuations $without$ hole doping, a crucial role of orbital
degree of freedom has been pointed out by Takimoto {\it et al.} \cite{Takimoto}.

Concerning Pu-115, it is still premature to draw a definitive conclusion
about the mechanism of superconductivity, but we notice that some
normal-state properties in PuMGa$_5$ seem to be dominated by
AFM spin fluctuations, e.g., the Curie-Weiss behavior in magnetic
susceptibility and electric resistivity in proportion to $T^{1.35}$
\cite{Sarrao}.
Thus, it may be natural to consider that superconductivity in
Pu-based compounds is also induced by AFM fluctuations.
However, several problems still exist, even if $d$-wave
superconductivity is confirmed in both Ce-115 and Pu-115
materials.
One question is, of course, why $T_{\rm c}$ is so high in Pu-115.
As is well known, due to difference in spatial extension of wavefunctions,
$5f$ electrons have intermediate nature between localized $4f$ and
itinerant $3d$ electrons.
Namely, energy scale of $5f$-electrons should be larger than that of
$4f$-electrons, leading to higher $T_{\rm c}$ in $5f$ electron systems
if we assume the same electronic mechanism for superconductivity.
However, this cannot be the whole story and the situation is
not so simple, since Pu$^{3+}$ ion includes five $f$-electrons,
in contrast to one $f$-electron in Ce$^{3+}$ ion.
Furthermore one has to address the question why U-115 does $not$
exhibit superconductivity.
If we follow the above scenario about energy scale, U-115 can be
superconducting with relatively high $T_{\rm c}$, but that is not the case.
Thus, it is $not$ sufficient to consider Pu-115 and U-115 as simple
analogues to Ce-115 with large energy scale,
based only on difference in itinerant nature between
$4f$- and $5f$-electrons.

In this Letter, in order to clarify those points,
we calculate energy band structures
and Fermi surfaces for PuCoGa$_5$ by applying a relativistic linear
augmented-plane-wave (RLAPW) method.
It is found that several sheets with large volume form cylindrical Fermi
surfaces, quite similar to CeMIn$_5$ (M=Ir and Co) \cite{Maehira1}.
The whole energy scale in the band structure of PuCoGa$_5$ is larger
than that of CeMIn$_5$, as naively expected.
On the other hand, for UCoGa$_5$, we have found only several small pocket
Fermi surfaces to show semi-metal like behavior \cite{Maehira2},
consistent with the fact that UMGa$_5$ is $not$ superconducting.
The similarity between Ce-115 and Pu-115 is understood by a simple
tight-binding model
constructed based on the $j$-$j$ coupling scheme \cite{Hotta}.
A remarkable fact is that Pu-115 can be regarded as a hole version of Ce-115.
Thus, we can conclude that 
both Ce-115 and Pu-115 have the same electronic origin for superconductivity,
suggesting that Pu-115 has higher $T_{\rm c}$ due to the combination
of the electron-hole conversion concept and the energy-scale discussion.


First let us briefly explain the RLAPW method.
Readers interested in the formalism can consult with Refs.~\cite{Hasegawa}.
When we calculate the electronic energy band structure of $4f$ and $5f$
compounds, in general, relativistic effects should be included,
since electrons near the heavy
nucleus must move with a high speed to keep their stationary motion.
In order to take into account major relativistic effects such as
the relativistic energy shifts, the relativistic screening effects,
and the spin-orbit interaction,
Loucks derived a relativistic augmented-plane-wave method
based on the Dirac one-electron wave equation \cite{Loucks}.
Several problems in his method have been improved by Hasegawa
and co-workers \cite{Hasegawa}.
The local density approximation is used for the exchange and correlation
potential and spatial shape of one-electron potential
is determined in the muffin-tin approximation.
Self-consistent calculations are performed by using the
lattice constants determined experimentally \cite{Sarrao}.

\begin{figure}
\includegraphics[width=1.0\linewidth]{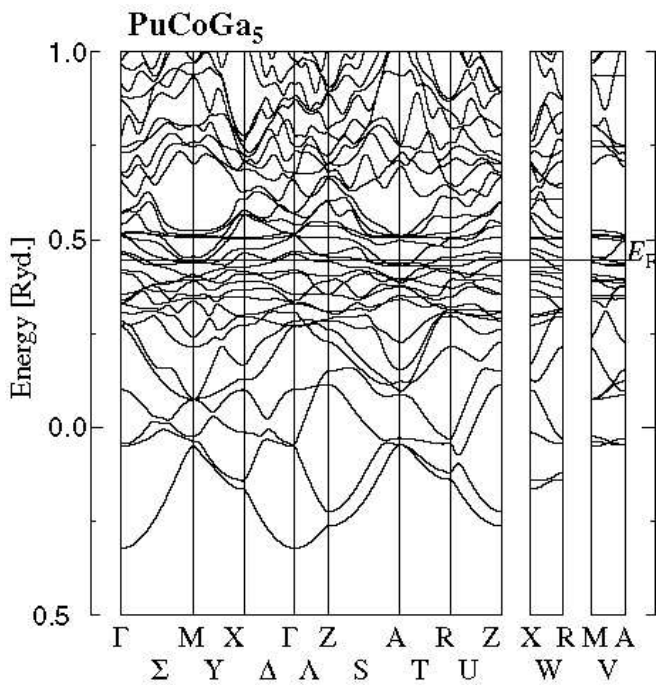}
\caption{Energy band structure for PuCoGa$_5$ obtained by
the RLAPW method.
$E_{\rm F}$ denotes the position of the Fermi level.
}
\end{figure}


In Fig.~1, we show the energy band structure for PuCoGa$_5$ along the symmetry
axes in the Brillouin zone in the range from $-$0.5Ryd. to 1.0Ryd.,
where Ryd. indicates Rydberg and 1Ryd.=13.6eV.
First note that in the vicinity of the Fermi level $E_{\rm F}$
located at 0.446 Ryd.,
there occurs hybridization between Pu 5$f$ and Ga 4$p$ states.
Above $E_{\rm F}$ near the M point, the flat 5$f$ bands split into two groups,
corresponding to the total angular momentum $j$=5/2 (lower bands) and
7/2 (upper bands).
The magnitude of the splitting $\Delta$ between the two groups is estimated
as $\Delta$(Pu)=1 eV, which is almost equal to the spin-orbit splitting
in the atomic $5f$ state of Pu.
Note that each Pu APW sphere contains about 5.2 electrons
in the $f$ state,
suggesting that valence of plutonium ion is Pu$^{3+}$,
consistent with experimental result \cite{Sarrao}.

By using the total density of states at $E_{\rm F}$, evaluated
as $N(E_{\rm F})$=97.3 states/Ryd.cell,
the theoretical specific heat coefficient
$\gamma_{\rm band}$ is estimated as 16.9 mJ/K$^2 \cdot$mol,
while the experimental electronic specific heat coefficient 
$\gamma_{\rm exp}$ is 77 mJ/K$^2 \cdot$mol \cite{Sarrao}.
If we define the enhancement factor for the electronic specific heat 
coefficient as $\lambda$=$\gamma_{\rm exp}/\gamma_{\rm band}$$-$1,
we obtain $\lambda$=3.6, which is smaller than $\lambda$=10
for CeCoIn$_5$ \cite{Maehira1}.
Note that the enhancement of $\lambda$ from unity is a measure
of electron correlation effect.
The moderate $\lambda$ in Pu-115 suggests that the correlation effect
in Pu-115 should be weak compared with Ce-115.
Since localized nature is stronger in $4f$ electrons, the correlation
effect is more significant in Ce-115.

\begin{figure}
\includegraphics[width=1.0\linewidth]{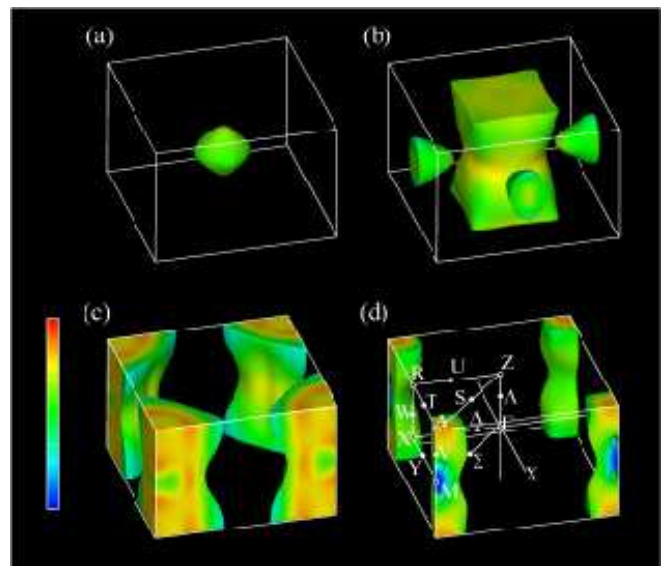}
\caption{Calculated Fermi surfaces of PuCoGa$_5$ for
(a) 15th band hole sheets, (b) 16th band hole sheets, 
(c) 17th band electron sheets, and 
(d) 18th band electron sheets.
Colors indicate the amount of $5f$ angular momentum character
on each Fermi surface sheet and 
red-shift indicate the increase of the $f$ charcter.
The center of the Brillouin zone is set at the $\Gamma$ point.}
\end{figure}

Now we discuss the Fermi surfaces of PuCoGa$_5$.
In Fig.~1, the lowest fourteen bands are fully occupied.
The next four bands are partially occupied, while higher bands are empty.
Then, as shown in Figs.~2(a)-(d), 15th, 16th, 17th, and 18th bands crossing
the Fermi level construct the hole or electron sheets of the Fermi surfaces,
summarized as follows:
(a) The Fermi surface from the 15th band includes one small hole sheet
centered at the $\Gamma$ point.
(b) The 16th band constructs a large cylindrical hole sheet centered at the
$\Gamma$ point, while two equivalent small hole sheets are centered at X points.
Depending on the energy resolution in the calculation, it is subtle whether
those two hole sheets touch each other or not,
but in any case, the quasi two-dimensional large hole sheet centered at the
$\Gamma$ point gives the main contribution. 
(c) The 17th band has a large cylindrical electron sheet centered
at the M point.
(d) The 18th band provides another cylindrical electron sheet centered
at the M point.

Let us consider the main Fermi surfaces from the viewpoints of
the Fermi-surface volume and $f$-electron admixture.
As for the Fermi surface constructed from 18th band, $f$-electrons are not
uniformly distributed on it, as expressed in color scale.
Around the A-point, $f$-electron admixture is large, while $p$-electron
gives a large contribuation around the M-point.
If we ignore three dimensionality and small-volume Fermi surfaces,
the main contributors are
the hole sheet from the 16th band centered at the $\Gamma$ point and
the electron sheet from the 17th band centered at the M point.

\begin{figure}
\includegraphics[width=1.0\linewidth]{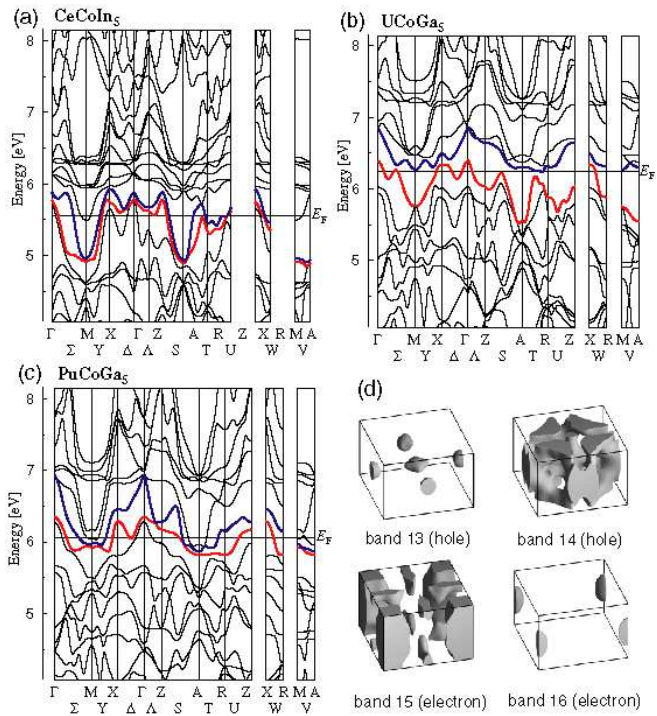}
\caption{Energy band structure around $E_{\rm F}$ for (a) CeCoIn$_5$,
(b) UCoGa$_5$, and (c) PuCoGa$_5$.
In each panel, red and blue curves indicate the upper and lower bands
to construct the hole and electron Fermi-surface sheets, respectively.
Namely, those are 14th (red) and 15th (blue) for CeCoIn$_5$,
15th (red) and 16th (blue) for UCoGa$_5$, and
16th (red) and 17th (blue) for PuCoGa$_5$.
(d) Calculated Fermi surfaces of CeCoIn$_5$.
}
\end{figure}

In order to gain deeper insight into electronic structure of Pu-115,
it is instructive to compare with the results for Ce-115 \cite{Maehira1}
and U-115 \cite{Maehira2}.
In Figs.~3 (a)-(c), we show the energy band structures around $E_{\rm F}$
for CeCoIn$_5$, UCoGa$_5$, and PuCoGa$_5$, respectively,
in the same energy scale.
Note that CeCoIn$_5$ becomes superconducting with $T_{\rm c}$=2.3K \cite{Ce115},
while UCoGa$_5$ is Pauli paramagnet \cite{U115}.
First, we can estimate $\Delta$(Ce)=0.4eV and $\Delta$(U)=0.8eV,
which are almost equal to the spin-orbit splittings in the atomic $4f$ and
$5f$ states for Ce and U, respectively.
As expected, we obtain $\Delta$(Ce)$<$$\Delta$(U)$\alt$$\Delta$(Pu).
As shown in the figure caption, the number to label red and blue curves
increase one by one in the order of CeCoIn$_5$, UCoGa$_5$, and PuCoGa$_5$,
corresponding to the increase in $f$-electron number by two per site.
Note that shapes of red and blue curves among three 115 compounds are similar
to one another, since overall band structure around the Fermi level is
always determined by hybridization between broad $p$-bands and
narrow $f$-bands for 115 compounds.
The center of gravity of the $j$=5/2 states
in CeCoIn$_5$ is about 0.4eV above $E_{\rm F}$, while the center of those
in PuCoGa$_5$ is slightly lower than $E_{\rm F}$.
Concerning UCoGa$_5$, the $j$=5/2 states seem to be just
at the Fermi energy.
This trend is consistent with the number of $f$-electrons
in each compounds.
The width of $j$=7/2 and 5/2 bands around at the M- or A-points becomes
broad in the order of CeCoIn$_5$, UCoGa$_5$, and PuCoGa$_5$,
consistent with the difference in $4f$- and $5f$-electron wavefunctions.

Here we emphasize that Ce-115 and Pu-115 exhibit large Fermi suafces,
as shown in Figs.~2 and 4(d).
In particular, we see a clear similarity between main Fermi surfaces of
CeCoIn$_5$ and PuCoGa$_5$, except for fine structures.
Considering only $f$-electron dominant Fermi surface with large volume,
we observe in common the large hole sheet centered at the $\Gamma$ point
and the large cylindrical electron sheet centered at the M point.
On the other hand, U-115 has small-pocket Fermi surfaces, as deduced from
Fig.~2(b) \cite{Maehira2}.
Namely, U-115 is considered as a semi-metal, which seems to be closely
related to the reason why U-115 does not exhibit superconductivity.
The origin of the semi-metallic behavior may be traced back
to slight overlap among the $j$=5/2 $f$-bands strongly hybridized
with the $p$-states from Ga ions.

\begin{figure}
\includegraphics[width=1.0\linewidth]{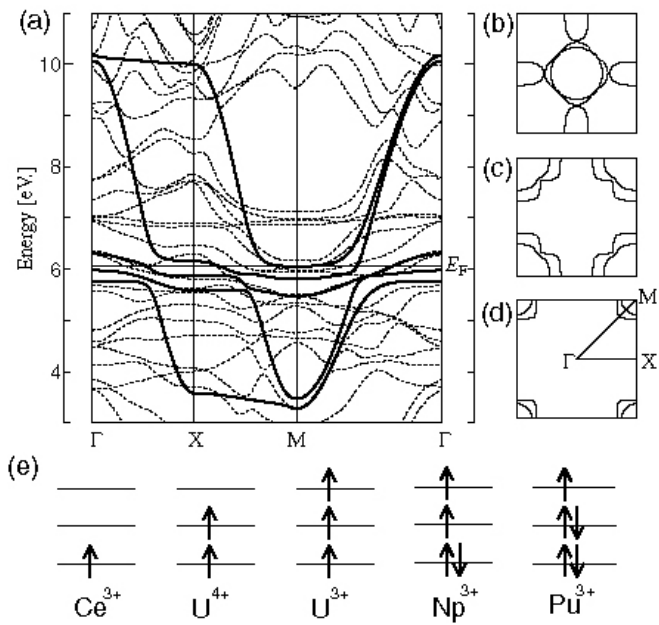}
\caption{(a)Energy band structures for PuCoGa$_5$ around $E_{\rm F}$
for the tight-binding model
(solid curves) and the RLAPW results(dashed curves).
Fermi-surface lines discussed here are
(b)16th band hole sheets, (c)17th band electron sheets, and
(d)18th band electron sheets.
Note that solid and broken curves denote the tight-binding and
RLAPW Fermi surfaces, respectively.
(e) Configurations for $f$-electrons accommodated in three Kramers doublets.
Up and down arrows denote pseudo-spin up and down, respectively.
}
\end{figure}

In order to understand the similarity in energy band structures and
Fermi surfaces between Ce-115 and Pu-115,
it is convenient to reanalyze the tight-binding model obtained based
on the $j$-$j$ coupling scheme \cite{Hotta}.
To consider the 115 systems, we include only $f$- and $p$-electrons
in the two-dimensional network composed of Ce and In (Pu and Ga) ions
\cite{Maehira1}.
Due to the lack of space, we skip the details of the model for $j$=5/2
sextet \cite{Maehira1,Hotta}, but the Hamiltonian $H$ is written as
$H$=$H_{\rm f}$+$H_{\rm p}$+$H_{\rm fp}$,
where $H_{\rm f}$, $H_{\rm p}$, and $H_{\rm fp}$ are, respectively,
$f$-electron hopping, $p$-electron hopping,
and $f$-$p$ hybridization terms, which are characterized by
the Slater integrals $(ff\sigma)$, $(pp\sigma)$, and $(fp\sigma)$,
respectively.
Note that crystalline electric field terms are simply ignored,
since those are much smaller than the energies considered here.

In Fig.~4(a), we show the direct comparison between
the RLAPW and tight-binding results for
$(ff\sigma)$=4500K, $(fp\sigma)$=6000K, and $(pp\sigma)$=18630K.
Top and bottom of the tight-binding bands are determined by comparison
with the RLAPW ones with significant amount of Ga $4p$ states.
The Fermi level for $H$ is determined so as to include five $f$ electrons.
First, overall features of the bands in the vicinity of $E_{\rm F}$ are well
reproduced by the mixture of broad $p$- and narrow $f$-bands.
Second, magnitude of parameters for PuCoGa$_5$ are large compared with
those for CeCoIn$_5$,
$(ff\sigma)$=4400K, $(fp\sigma)$=5360K, and $(pp\sigma)$=5730K
\cite{Maehira1}.
Note that the difference in $(pp\sigma)$ between PuCoGa$_5$ and CeCoIn$_5$
is mainly due to the difference of Ga $4p$ and In $5p$ electronic states.
Then, we conclude that $5f$ electrons are more itinerant than $4f$ ones from
the present results for tight-binding fitting.
As shown in Figs.~4(b)-(d), the main Fermi surfaces are well reproduced
by the tight-binding model.
Good agreements between RLAPW and tight-binding results indicate
the validity of the $j$-$j$ coupling scheme for PuCoGa$_5$.

In the tight-binding model constructed based on the $j$-$j$ coupling scheme,
we are allowed to consider the $f^n$ configuration with $n$$>$2
by accommodating $n$ electrons among three Kramers doublets.
In Fig.~4(e), we show several $f$-electron configurations.
First of all, we note the electron-hole conversion relation between
Ce$^{3+}$ and Pu$^{3+}$ ions \cite{Hotta}.
Thus, Pu-115 can be considered as a hole version of Ce-115 and
this is the very reason why common Fermi surfaces are observed.

Note that UMGa$_5$ has been found to be AFM metal for M=Ni, Pd, and Pt,
while Pauli paramagnetic for M=Fe, Co, and Rh \cite{U115}.
Thus, the present band-structure calculation assuming
the paramagnetic phase
is consistent with the experimental result for UCoGa$_5$.
On the other hand, a hint to understand AFM metallic behavior for
M=Ni, Pd, and Pt may be found in the local spin structure.
Although it is difficult to determine the exact valence
of uranium ion, it should be between U$^{4+}$ and U$^{3+}$.
As shown in Fig.~4(e), for U$^{4+}$ (U$^{3+}$) ion,
local spin $S$=1 (3/2) may be formed due to the Hund's
rule coupling and thus, the AFM phase will be favored.
However, it is still an open problem to explain the metallic behavior
in the AFM phase as well as the difference in the AFM spin structure
between UPtGa$_5$ and UNiGa$_5$ \cite{U115}.

Finally, we provide one short comment on Np-115.
For Np$^{3+}$ ion, as shown in Fig.~4(e), we can regard it as a hole version
of U$^{4+}$ and thus, Np-115 may not exhibit superconductivity, but
antiferromagnetism or paramagnetism, as an analogue of U-115.

In summary, we have performed the band-structure calculation for PuCoGa$_5$
and obtained the Fermi surfaces similar to Ce-115 materials.
This similarity can be understood by the electron-hole conversion picture
based on the $j$-$j$ coupling scheme.
We believe that high $T_{\rm c}$ in PuCoGa$_5$ can be understood 
by combining our electron-hole picture
with the energy-scale difference in $4f$- and $5f$-electrons.

We thank Y. \=Onuki, J. L. Sarrao, T. Takimoto, F. Wastin,
and H. Yamagami for discussions.
T. H. and K. U. are supported by the Grant-in-Aid
for Scientific Research from Japan Society for the Promotion of Science.


\end{document}